\documentclass[aps,prd,superscriptaddress,nofootinbib,amsmath,amsfonts,preprintnumbers,groupedaddress,showpacs,10pt,english]{revtex4}
\usepackage{amsmath}
\usepackage{amssymb}
\usepackage{babel}
\usepackage{wrapfig}
\usepackage{cancel}
\usepackage{bigints}

\newcommand{\e}{\mathrm{e}}
\usepackage{relsize,exscale}
\makeatletter

\usepackage{array,multirow,graphicx}
\usepackage{dcolumn}
\usepackage{newlfont}
\usepackage{bm}
\usepackage[colorlinks,citecolor=blue,urlcolor=blue,linkcolor=blue]{hyperref}
\usepackage[figtopcap]{subfigure}
\usepackage{color}

\allowdisplaybreaks[4]

\usepackage{scalerel}
\usepackage{tikz}
\usetikzlibrary{svg.path}
\definecolor{orcidlogocol}{HTML}{A6CE39}
\tikzset{
  orcidlogo/.pic={
    \fill[orcidlogocol] svg{M256,128c0,70.7-57.3,128-128,128C57.3,256,0,198.7,0,128C0,57.3,57.3,0,128,0C198.7,0,256,57.3,256,128z};
    \fill[white] svg{M86.3,186.2H70.9V79.1h15.4v48.4V186.2z}
                 svg{M108.9,79.1h41.6c39.6,0,57,28.3,57,53.6c0,27.5-21.5,53.6-56.8,53.6h-41.8V79.1z M124.3,172.4h24.5c34.9,0,42.9-26.5,42.9-39.7c0-21.5-13.7-39.7-43.7-39.7h-23.7V172.4z}
                 svg{M88.7,56.8c0,5.5-4.5,10.1-10.1,10.1c-5.6,0-10.1-4.6-10.1-10.1c0-5.6,4.5-10.1,10.1-10.1C84.2,46.7,88.7,51.3,88.7,56.8z};}}
\newcommand\orcid[1]{\href{https://orcid.org/#1}{\mbox{\scalerel*{
\begin{tikzpicture}[yscale=-1,transform shape]
\pic{orcidlogo};
\end{tikzpicture}
}{|}}}}
\begin{document}

\tolerance=5000

\title{The key role of Lagrangian multiplier in mimetic gravitational theory in the frame of  { isotropic} compact star}

\author{G.~G.~L.~Nashed~\orcid{0000-0001-5544-1119}}
\email{nashed@bue.edu.eg}
\affiliation {Centre for Theoretical Physics, The British University in Egypt, P.O. Box
43, El Sherouk City, Cairo 11837, Egypt\\
Center for Space Research, North-West University, Potchefstroom 2520, South Africa}

\date{\today}

\begin{abstract}
Recently, the mimetic gravitational theory has gained much attention in the frame of cosmology as well as in the domain of astrophysics. In this study, we show that in the frame of mimetic gravitation theory we are not able to derive an isotropic model. As a result, our focus shifts towards combining mimetic gravitational theory with the Lagrangian multiplier. The field equations of a static isotropic gravitational system that controls the geometry and dynamics of star structure are studied in the frame of mimetic theory coupled with a Lagrangian multiplier using a non-linear equation of state.
An energy density is assumed from where all the other unknowns are fixed and a new isotropic model is derived. The physical analysis of this model is studied from different viewpoints and consistent results compatible with a realistic isotropic star are investigated  analytically and graphically. Ultimately, we demonstrate the stability of the model in question by employing the adiabatic index technique.
\end{abstract}


\maketitle
\section{Introduction}\label{S1}

Recently, conclusive evidence developed proposing that Einstein's theory of general relativity (GR), is needy to be modified.The justifications for this observation stem from the challenges in renormalizing general relativity and the uncertain behavior exhibited in high gravity regions, such as the exterior of black holes and neutron stars.  Additionally, the confirmed accelerated expansion of our universe can not be explained by GR.  To be able to use GR to explain the phenomena of accelerated expansion of our universe an assumption of the presence of exotic matter fields like dark matter and dark energy must be taken into account. Up to date no experimental support for these was imminent. Another method is to try to amend Einstein's GR somehow so that we keep its basic gains.  Despite these issues in  GR, it still submits success in the solar system. The success of the Event Horizon Telescope in capturing the image of a black hole's shadow in M87 \cite{EventHorizonTelescope:2019dse} and the advancements made in detecting gravitational waves \cite{LIGOScientific:2016aoc} have significantly bolstered the position of General Relativity as the preeminent gravitational theory compared to other theories of gravity. However, aspects that GR's have not clarified must also face. We advocate the perspective that a modification of the governing field equations within the geometric sector holds the key to addressing these unresolved concerns.

Modified gravitational theories have made significant progress and performance in explaining some of the unsolved issues in GR. { There are many modified theories of gravity that can overcome the shortcomings of GR. One of the modification of GR is to add a scalar field. In the frame of a scalar field coupled with Ricci scalar a static neutron star perspective using two types of cosmological inflationary attractor theories, i.e., the induced inflationary attractors and the quadratic inflationary attractors are investigated \cite{Oikonomou:2023lnh}.
 Among these modification is the $f(R)$ gravitational theory \citep[see for exmaple][]{Nojiri:2019dqc,Odintsov:2019ofr}.  In the frame of $f(R)$  a study of the neutron star phenomenology of $R^p$ attractor theories in the Einstein frame have been analyzed \cite{Oikonomou:2023dgu}. In this study we are interested in another  modification of  GR, i.e., we are focused on  the gravitational
mimetic theory which  has been suggested as a fresh approach to studying the problem of dark matter.\cite{Chamseddine:2013kea}.} This concerns the introduction of a mimetic scalar field denoted as $\eta$, which, despite lacking dynamics in its construction, plays a crucial role in imparting dynamism to the longitudinal degree of freedom within the gravitational system.  The gravitational system's dynamic longitudinal degree serves as an analogue to pressureless dark matter, mimicking its properties  \cite{Chamseddine:2013kea}. The problem of cosmological singularities \cite{Chamseddine:2016uef}  and the singularity at the core of a black hole  \cite{Chamseddine:2016ktu} can be effectively tackled through the altered variant of mimetic gravitational theory.  Furthermore, the gravitational theory of mimetic has substantiated that the propagation of gravitational waves at the speed of light aligns in perfect accordance with the discoveries made from the event GW170817 and the corresponding optical observations  \cite{Casalino:2018tcd,Casalino:2018wnc,Sherafati:2021iir}. Furthermore, it has been demonstrated that mimetic theory can explore the coherent rotational patterns observed in spiral galaxies without relying on the existence of dark matter particles \cite{Vagnozzi:2017ilo,Sheykhi:2019gvk}. Lately, there has been a significant surge of enthusiasm surrounding the cosmological framework due to the emergence of the mimetic theory \cite{Chamseddine:2014vna,Baffou:2017pao,Dutta:2017fjw,Sheykhi:2018ffj,Abbassi:2018ywq,Matsumoto:2016rsa,Sebastiani:2016ras,Sadeghnezhad:2017hmr,Gorji:2019ttx,
Gorji:2018okn,Bouhmadi-Lopez:2017lbx,Gorji:2017cai,Chamseddine:2019bcn,Russ:2021ede,deCesare:2020swb,Cardenas:2020srs,HosseiniMansoori:2020mxj,Arroja:2017msd} and  black
holes physics  \cite{Deruelle:2014zza,Myrzakulov:2015sea,Myrzakulov:2015kda,Ganz:2018mqi,Chen:2017ify,Nashed:2018aai,BenAchour:2017ivq,Brahma:2018dwx,Zheng:2017qfs,Shen:2019nyp,Nashed:2018qag,Chamseddine:2019pux,Gorji:2020ten,Sheykhi:2020fqf,Sheykhi:2020fqf,Nashed:2021ctg,Chamseddine:2021xhw,Zheng:2018cuc,Bakhtiarizadeh:2021pyo,Nashed:2021pkc}.
 The theory has further extended its scope to include $f(R)$ mimetic gravity, incorporating additional insights and explanations \cite{Nojiri:2014zqa,Odintsov:2015wwp,Oikonomou:2016pkp,Oikonomou:2016fxb,Oikonomou:2015lgy,Myrzakulov:2016hrx,Odintsov:2015cwa,Odintsov:2016imq,Odintsov:2016oyz,
Nojiri:2017ygt,Odintsov:2018ggm,Bhattacharjee:2021kar,Kaczmarek:2021psy,Chen:2020zzs} and mimetic Gauss-Bonnet gravity \cite{Astashenok:2015haa,Oikonomou:2015qha,Zhong:2016mfv,Zhong:2018tqn,Paul:2020bje}. Specifically, a comprehensive framework combining early inflation and late-time acceleration within the context of mimetic $f(R)$ gravity was formulated \cite{Nojiri:2016vhu}. It has been stressed that within the context of mimetic $f(R)$ gravity, the period of inflation can be identified \cite{Nojiri:2016vhu}.

Nowadays, the mimetic theory is one of the most compelling theories of gravity, which without introducing any additional field of matter, represents the dark side of the universe which is represented as a geometric effect.  The observations ensure that approximately 26$\%$ of the energy
content of the universe is related to the dark matter sector,
while approximately 69$\%$ construct the dark energy \cite{Yang:2019fjt}. Many facts ensure the presence of dark matter and
dark energy \cite{Seljak:2006bg}. Dark energy, which has gained significance in recent times, is believed to be a smooth element characterized by negative pressure. It possesses an anti-gravity characteristic and is actively propelling the universe, playing a key role in the accelerated expansion of the cosmos \cite{Oks:2021hef}. Dark matter performs two crucial functions in the development of the universe: Firstly, it provides the necessary gravitational force for the rotation of spiral galaxies and galaxy clusters. Secondly, it plays a significant role as an essential component in the amplification of disturbances and the formation of structures during the early stages of the universe.  As a result, dark matter begins to condense into an intricate system of dark matter halos, whereas regular matter undergoes collapse due to photon radiation, eventually settling into the well-formed potential of dark matter.  In the absence of dark matter, the process of galaxy formation in the universe would be significantly more extensive than what has been observed.

The structure of the current investigation is outlined as: In Sec.~\ref{S2}, we introduce the fundamental principles of the mimetic theory combined with the Lagrangian multiplier. In Sec.~\ref{S3}, we list the necessary conditions that must be obeyed by any realistic isotropic model. Also,  in Sec.n~\ref{S3} we show that the model under consideration satisfies all the necessary conditions, analytically and graphically,  that must possess by any realistic isotropic star. In Section~\ref{S4} we study the stability of the model presented in this study using the adiabatic index. Final Section is devoted to discussing the main results of the present study.

\section{Isotropic  solution in the mimetic theory combined with  the Lagrange multiplier}\label{S2}

What is  called ``dark matter of mimetic" was delivered   in the scientific society in \cite{Chamseddine:2014vna}.
Despite of this  mimetic theories had already been discussed in \cite{Lim:2010yk,Gao:2010gj,Nashed:2022yfc,Nashed:2021hgn,Nashed:2021pkc,Nashed:2021ctg,Nashed:2018urj,Nashed:2018qag,Nashed:2018aai,Capozziello:2010uv,Sebastiani:2016ras}.

According to the theory of mimetic, the
metric $g_{\alpha \beta}$, which is a physical one, is linked to the metric  $\bar{g}_{\alpha \beta}$,  which is an auxiliary metric, and to the
mimetic scalar field $\eta$ by the  conformal transformation:
\begin{equation}
\label{trans1}
g_{\alpha\beta}=- \left(\bar{g}^{\mu \nu} \partial_\mu \eta \partial_\nu \eta \right) \bar{g}_{\alpha\beta}\,,
\end{equation}
 where the metric $\bar{g}^{\alpha \beta}$ undergoes conformal transformations, specifically $\bar{g}^{\alpha \beta} \to \Omega \bar{g}^{\alpha \beta}$, the metric $g^{\alpha \beta}$ remains invariant, meaning that it remains unchanged.

In the present  study, we will  coin  the { mimetic-like gravity} coupled with   the Lagrange multiplier.
The action of the gravity model resembling the mimetic theory, in conjunction with the Lagrange multiplier $\lambda$ and the function $\omega$, is described as follows:
has the form\footnote{It was shown in \cite{Nojiri:2022cah} that the function $\omega$ is  necessary in the construction of the mimetic theory  which allows us to achieve the geometric characteristics of a black hole, including the presence of one or more horizons.} \cite{Nojiri:2014zqa}:
\begin{equation}
\label{actionmimeticfraction}
S=\int \mathrm{d}x^4\sqrt{-g}\left\{ R
 +\lambda \left(g^{\mu \nu}\omega \partial_{\mu}\eta\partial_{\nu}\eta
+1\right)\right \}+L_\mathrm{matt}\, ,
\end{equation}
where $L_\mathrm{matt}$ is the Lagrangian of the matter field and $\eta$ is the mimetic scalar field.
%
%
The field equations can be obtained by taking the variation of the action (\ref{actionmimeticfraction}) with respect to the metric tensor $g_{\mu \nu}$, resulting in the following expressions:
\begin{align}
\label{aeden}
& 0=R_{\mu \nu}-\frac{1}{2}g_{\mu \nu}R
+\frac{1}{2}g_{\mu \nu}\left\{\lambda \left( g^{\rho
\sigma}\omega\partial_{\rho}\eta\partial_{\sigma}\eta+1\right) \right\}-\lambda
\partial_{\mu}\eta \partial_{\nu}\eta +\frac{1}{2}T_{\mu \nu} \, ,
\end{align}
where $T_{\mu \nu}$ is the energy-momentum tensor corresponding to the matter field.
Additionally, differentiating the action (\ref{actionmimeticfraction}) with respect to the mimetic  field $\eta$ yields:
\begin{equation}\label{scalvar}
2\nabla^{\mu} (\lambda \omega\, \partial_{\mu}\eta)=0\, .
\end{equation}
When the action  (\ref{actionmimeticfraction})  is varied with respect to the Lagrange multiplier $\lambda$, the following outcome is obtained:
\begin{equation}\label{lambdavar}
g^{\rho \sigma}\omega \partial_{\rho}\eta \partial_{\sigma}\eta=-1\, .
\end{equation}
It is the time to apply the field equations (\ref{aeden}) and (\ref{scalvar}), using the constrains of Eq. (\ref{lambdavar}),  to the following  spherically symmetric spacetime

\begin{equation}
\label{met}
ds^2=f(r)dt^2-\frac{dr^2}{f_1(r)}-r^2\left(d\theta^2+r^2\sin^2\theta d\phi^2\right)\,,
\end{equation}
 to derive an interior solution. The functions $f(r)$ and $f_1(r)$ mentioned here are unfamiliar functions that will be determined by solving the set of field equations.

Furthermore, we assume that $\eta$ is solely dependent on   $r$.
Using Eqs.~(\ref{aeden}) and (\ref{scalvar}) to the spacetime (\ref{met}), yields the following set of differential equations.
\noindent
The $(t,t)$-component of the field equation (\ref{aeden}) is:
\begin{align}
\label{eqtt}
\rho(r)=&\frac{1 - f_1 - rf'_1}{r^2} \,,
\end{align}
the field equation (\ref{aeden}) can be expressed as the $(r,r)$-component.
\begin{align}
\label{eqrr}
p(r)=&\frac{ f_1 f'r-f + f f_1-\lambda \omega(r)\eta'^2 ff_1  r^2}{r^2f}\,,
\end{align}
the components of the field equation (\ref{aeden}) in terms of $(\theta,\theta)$ and $(\phi,\phi)$ have the following structure:
\begin{align}
\label{fe}
p(r)=& \frac {2\,f_1 f''f  r- f'^{2}f_1  r+f
   \left( 2\,f_1  + f'_1r \right) f' +2\,f'_1 f^{2}}{4f^{2}r}
\,,
\end{align}
and the field equation (\ref{scalvar}) takes the form:
\begin{equation}
\label{chr}
0=2 \lambda'\omega
 f r+ \left[ \omega' f  r+\omega
 \left(  f' r+4\,f  \right)  \right] \lambda
 \,,
\end{equation}
where $f\,\equiv f(r)$, $f_1\,\equiv f_1(r)$, $\omega\,\equiv \omega(r)$,  $\lambda\,\equiv \lambda(r)$, $f'=\frac{df}{dr}$, $f'_1=\frac{df_1}{dr}$, $\eta'=\frac{d\eta}{dr}$, $\omega'=\frac{d\omega}{dr}$, and $\lambda'=\frac{d\lambda}{dr}$.
Furthermore, we make the assumption that the energy-momentum tensor of the isotropic fluid can be represented in the following manner:
\begin{eqnarray}
{T}_{\mu \nu} = (\rho+p)u_\mu u_\nu+p g_{\mu \nu}\,,
\end{eqnarray}
In this case, $\rho$ represents the energy density, $p$ denotes the pressure, and $u^\mu$ is a timelike vector defined as $u^\mu = [1, 0, 0, 0]$.In this particular investigation, we consider the matter content to be characterized by the energy density $\rho$ and the pressure $p$ respectively.\footnote{In the entirety of this research, we will utilize geometrized units where the constants $G$ and $c$ are set to 1.}.

Assuming the fluid under consideration has a perfect fluid verifying the equation of state (EoS),
$p=p\left(\rho\right)$.
Using the conservation law of  matter gives:
\begin{align}
\label{FRN2}
0 = \nabla^\mu \, T_{\mu r} =2f\frac{dp}{dr}+f' \left( \rho + p \right)  \, .
\end{align}
In this particular context, we make an assumption that the energy density and pressure of the system exhibit variations based on the radial coordinate.

If the form of EoS $\rho=\rho(p)$ is presented, then Eq.~(\ref{FRN2}) yield:
\begin{align}
\label{FRN3}
\frac{1}{2} \ln f = - \int^r dr \frac{\frac{dp}{dr}}{\rho + p}
= - \int^{p(r)}\frac{dp}{\rho(p) + p} \, .
\end{align}
In the interior of the  star,  Eq.~(\ref{FRN3}) can be used but in the exterior, Eq.~(\ref{FRN3}) cannot be used.
Nevertheless, it is possible to assume that both $f(r)$ and its derivative $f'(r)$ exhibit continuity at the surface of the star.

Considering the count of unknown functions and independent equations, we find a total of six unknown functions within the compact star. To address this, we will employ the constraint given by Eq. (\ref{lambdavar}), which states that $\eta=\frac{1}{\sqrt{-\omega f_1}}$. These unknown functions include two metric potentials, namely $f(r)$ and $f_1(r)$ in (\ref{met}), as well as the Lagrangian multiplier, the function $\omega$, the energy density $\rho$, and the pressure $p$ present in the action (\ref{actionmimeticfraction}). However, we have 3 independent equations, that are, the three components, Eqs.~(\ref{eqtt}), (\ref{eqrr}), and (\ref{fe})
of the field equation (\ref{aeden}) of the mimetic  equations, an equation of state $\rho=\rho(p)$,
and the conservation law (\ref{FRN2}). As we mentioned above, the scalar field equation (\ref{scalvar}) can be obtained from the field equations (\ref{aeden}) corresponding
to the mimetic equation and the conservation law $\nabla_\mu T^{\mu\nu}=0$ of the matter, and therefore the scalar field equation (\ref{scalvar}) is not independent.
Hence, we have $6-5=1$ undetermined function remaining from the equations. To address this remaining aspect, we opt for the specific configuration of energy density $\rho=\rho(r)$ within the compact celestial object.
Moreover, outside the star, we have $\rho=p=0$, and the unknown functions, $f(r)$, $f_1(r)$ and $\eta$ (or $\omega(\eta)$).
We have 3 independent equations, that are, the three components, Eqs.~(\ref{eqtt}), (\ref{eqrr}), and (\ref{fe}).
Thus there are $5-4=1$ function undetermined from the equations, again.

If we consider a compact star like neutron star, one usually consider the  EoS as:
\begin{enumerate}
\item Energy-polytrope
\begin{align}
\label{polytrope}
p = k \rho^{1 + \frac{1}{s}}\,,
\end{align}
where  $k$ and $s$ are constants.
It is well known that for  neutron star, $s$ lies in the interval  $s\in[0.5\,,\,1]$.
\item Mass-polytrope
\begin{align}
\label{MassPolytropicEOS}
\rho = \rho_m + s_1 p \, ,\qquad \qquad p = m_m \rho_m^{1+\frac{1}{s_m}} \, ,
\end{align}
with $\rho_m$  being the rest mass energy density and $m_m$, $s_1$, and $s_m$ are constants.
\end{enumerate}

Now, let us turn our attention to the study of the energy-polytrope case.
 Then EoS (\ref{polytrope})  can be rewritten   as:
\begin{align}
\label{polytrope2}
\rho = \tilde k p^{({1 + \frac{1}{\tilde s}})}\, , \quad
\tilde k \equiv k^{-\frac{1}{1+\frac{1}{s}}} \, , \quad
\tilde s \equiv \frac{1}{\frac{1}{1+\frac{1}{s}} - 1}
= - 1 - s \, .
\end{align}
 Eq.~(\ref{FRN3}) can take the  form:
\begin{align}
\label{FRN3p1B}
\frac{1}{2} \ln f = - \int^{p(r)}\frac{dp}{\tilde k p^{1 + \frac{1}{\tilde s}} + p}
= \frac{c_1}{2} + \tilde s \ln \left(1+{\tilde k}^{-1}p^{-\frac{1}{\tilde s}} \right)
= \frac{c_1}{2} - \left(1+s\right) \ln \left(1+ k \rho^\frac{1}{s} \right) \, ,
\end{align}
where $c_1$ is a constant of  integration.
Using the same method of polytrope we get for mass-polytrope (\ref{MassPolytropicEOS}) the function $f$ as:
\begin{align}
\label{masspolytope}
\frac{1}{2} \ln f = \frac{\tilde c}{2} + \ln \left( 1 - k_m \rho_m{}^\frac{1}{s_m}\right) \,,
\end{align}
where $\tilde c$ is a constant of integration.

To provide an illustrative example, taking into account one of the mentioned equations of state, we can make an assumption about the profile of $\rho=\rho(r)$ as follows.
\begin{align}
\label{anz1}
\rho=\left\{ \begin{array}{cc}
\rho_0 \left( 1 - \frac{r}{{R}} \right) & \ \mbox{when}\ r<R \\
0 & \ \ \mbox{when}\ r\geq R
\end{array} \right.   \, .
\end{align}
In this scenario, $\rho_0$ represents a fixed value expressing the energy concentration at the core of the condensed celestial object, whereas $R$ symbolizes an additional fixed value denoting the size of the compact star's outer boundary. As clear from Eq. (\ref{anz1}), the energy density $\rho$ vanishes at the surface $r=R$.
By using the energy-polytrope EoS (\ref{polytrope}) or the mass-polytrope EoS (\ref{MassPolytropicEOS}), we find that the pressure $p$ also vanishes at the surface.
We have introduced the mass parameter $M$ as a constant value, which is associated with the mass of the compact star. This parameter is defined specifically for the polytropic equation of state (EoS) as follows:
\begin{align}
\label{MRs}
M =4\pi \int_0^{R} y^2 \rho(y) dy
= \frac{\pi \rho_0 R^3}{3}\, .
\end{align}

Then Eq.~(\ref{FRN3p1B}) gives,
\begin{align}
\label{FRN3p1BC1}
f = \frac{\e^{c_1}}{ \left( 1 + k \rho_0 \left( 1 - \frac{r}{{R}} \right) \right)^4} \,.
\end{align}
Using Eq. (\ref{anz1}) in (\ref{eqtt}) we get:
\begin{align}
\label{f1}
f_1 =1-\frac{8\pi r^2}{3}+\frac{2r^3}{R}\,.
\end{align}
The Lagrangian multiplier of the above model has the form
\begin{align}
\label{FRN3p1BC12}
\lambda(r)=&\frac {c_2\, \left( R+kR-kr \right) ^{2}}{{r}^{5/2}}\,.
\end{align}
To finalize the determination of the remaining unknowns we assume, for simplicity,  the form of the function $\omega=c_3 r$ and the mimetic scalar field becomes:
\begin{align}
\label{FRN3p1BC13}
\eta(r)=&\sqrt{c_3 r\left(\frac{8\pi r^2}{3}-\frac{2r^3}{R}-1\right)}\,.
\end{align}
To finalize this section we stress on the fact that  if we follow the same procedure and put $\lambda(r)=0$ in Eqs. (\ref{eqtt}), (\ref{eqrr}), and (\ref{fe}) we get a system that we cannot  derive from it an isotropic model.
\section{Necessary conditions for a real physical  star }\label{S3}

Any physical reliable isotropic star model must obey the below conditions in the interior configurations:\vspace{0.1cm}\\
$\bullet$ It is essential for the metric potentials ($g_{tt}$ and $g_rr$) to provide precise explanations for the density and momentum components   by establishing unambiguous definitions and displaying consistent patterns within the core of the star and its interior.\vspace{0.1cm}\\
$\bullet$  Within the interior of the star, it is required that the energy density $\rho$ maintains a non-negative value, that is, $\rho \geq 0$. Additionally, the energy density has a finite positive value at the central region of the star and shows a decreasing pattern as it extends towards the surface, characterized by the condition $\frac{d\rho}{dr}\leq 0$.\vspace{0.1cm}\\
$\bullet$  Inside the fluid configuration, it is necessary for the pressure $p$ to be positive or zero ($p \geq 0$). Furthermore, within the interior of the star, it is expected that the pressure decreases with respect to the radial coordinate, as indicated by the condition $\frac{dp}{dr} < 0$. On the outermost boundary of the star, specifically at the surface where the distance from the center is denoted as $r = R$, the pressure $p$ should be precisely zero. This implies that there is no pressure exerted at the star's outer edge.\vspace{0.1cm}\\
$\bullet$  The energy conditions of an isotropic fluid sphere are given by: \vspace{0.1cm}\\
(i) The null energy condition (NEC) implies that the energy density $\rho$ must be greater than zero.\vspace{0.1cm}\\
(ii) According to the weak energy condition (WEC), the sum of the pressure $p$ and the energy density $\rho$ must be greater than zero, i.e., $p + \rho > 0$.\vspace{0.1cm}\\
(iii) In accordance with the strong energy condition (SEC), the sum of the energy density $\rho$ and three times the pressure $p$ must be greater than zero, i.e., $\rho + 3p > 0$.\vspace{0.1cm}\\
$\bullet$Furthermore, to ensure a realistic model, the causality condition must be satisfied within the interior of the star. This condition imposes a restriction on the speed of sound, requiring it to be less than 1. In this context, assuming the speed of light $c$ is equal to 1, the condition can be expressed as $\frac{dp}{dr} > 0$ and $1 > \frac{dp}{dr}$.\vspace{0.1cm}\\
$\bullet$  Finally,  the adiabatic index must has a value more than $\frac{4}{3}$.\vspace{0.1cm}\\

It is our purpose to study the above conditions on the isotropic model and see if it is real model or not.

\section{The physical characteristics of the model}\label{S4}
To determine whether the model described by Eqs. (\ref{FRN3p1B}), (\ref{anz1}), and (\ref{f1}) corresponds to a realistic stellar structure, we will examine the following aspects:
\subsection{Non singular model}
\begin{figure}
\centering
\subfigure[~Metric potentials, $f(r)$ and $f_1(r)$ given by Eq. (\ref{FRN3p1BC1}) and (\ref{f1})]{\label{fig:pot2}\includegraphics[scale=.3]{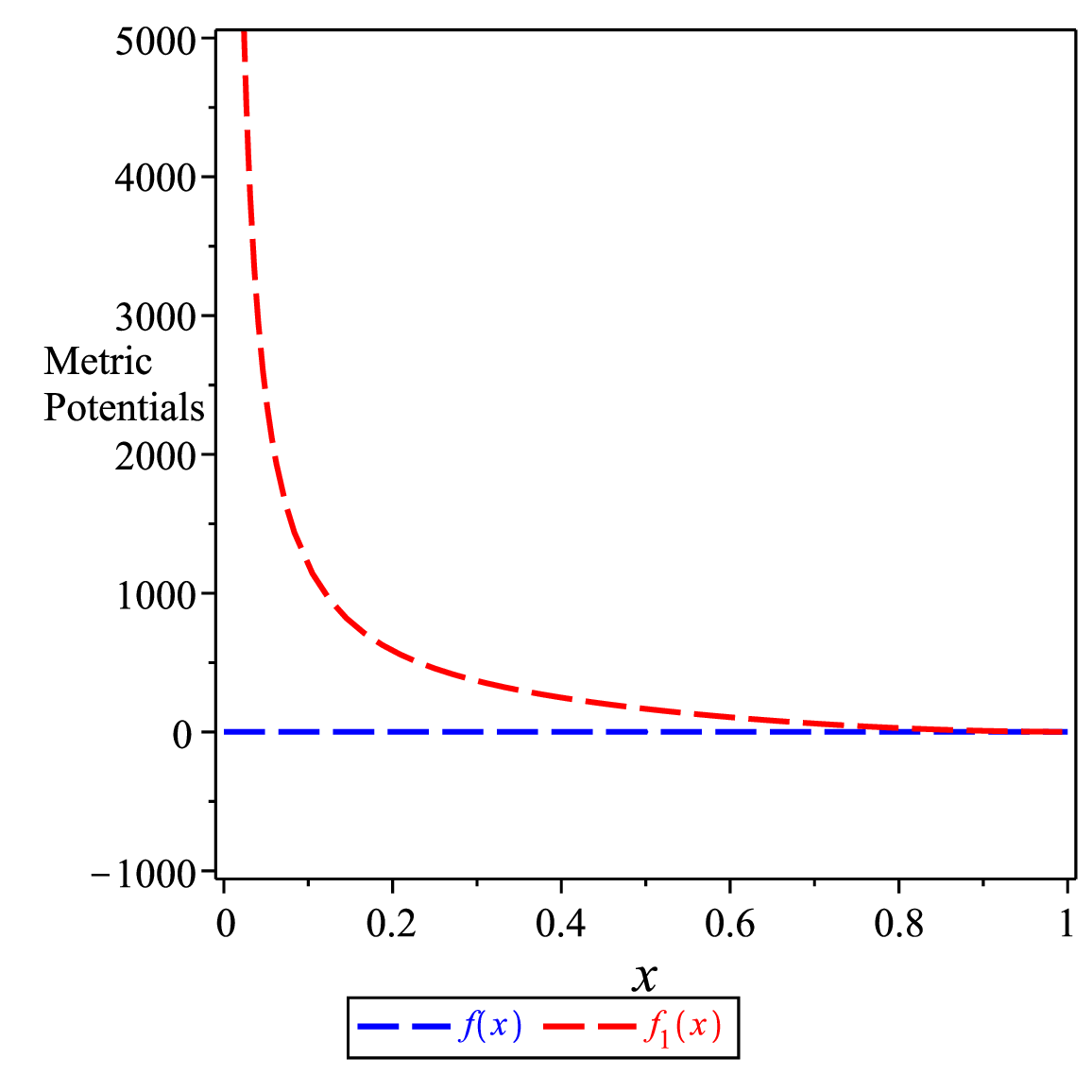}}
\subfigure[~Density]{\label{fig:density}\includegraphics[scale=0.3]{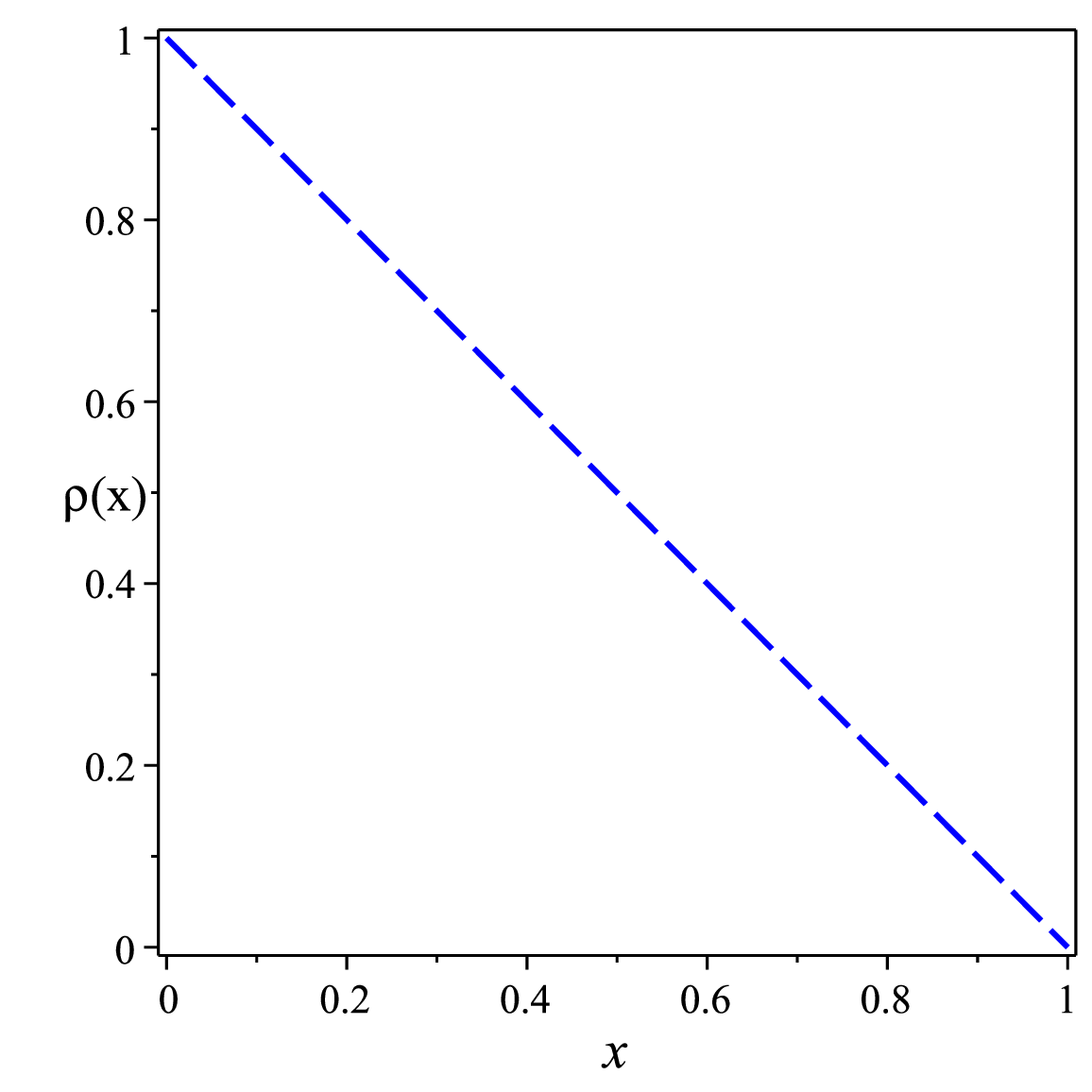}}
\subfigure[~Pressure]{\label{fig:pressure1}\includegraphics[scale=.3]{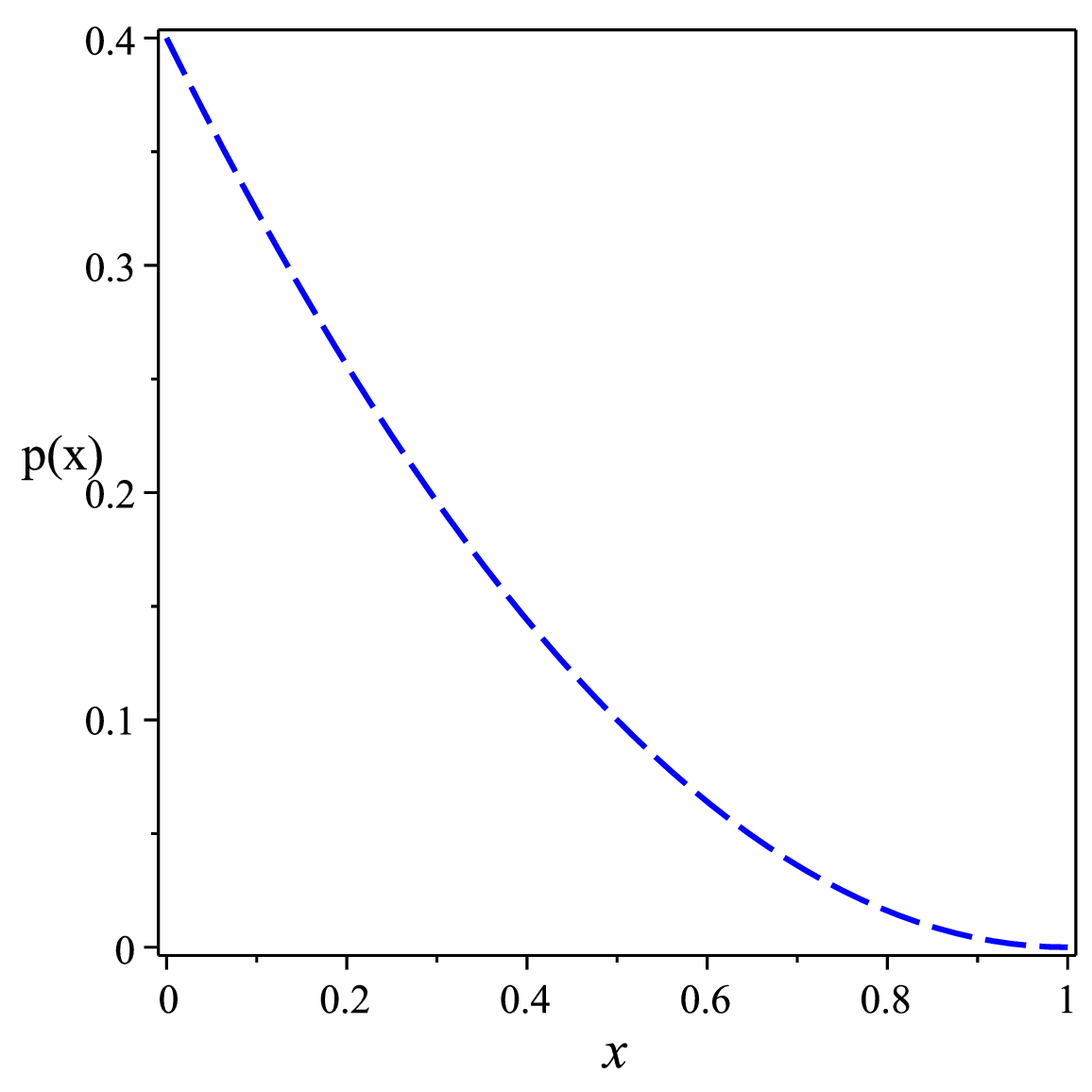}}
\caption[figtopcap]{\small{{A visual representation is provided, illustrating the relationship between the metric potentials (\ref{FRN3p1BC1}) and (\ref{f1}), in comparison to the dimensionless x; \subref{fig:density} the profile of density; and \subref{fig:pressure1} profile of pressure. We have put $\rho_0=1$ and $K=0.4$.}}}
\label{Fig:1}
\end{figure}

i- The components of the metric potentials $g_{tt}$ and $g_{rr}$ fulfill the following conditions:\footnote{We will rewrite all the physical quantities like metric potentials, density and pressure in terms of the dimensionless quantity $x$ where $x=\frac{r}{R}$.},
\begin{align}\label{sing}
f(x\rightarrow 0)=\frac{e^c}{(1+k\rho_c)^4}\,\qquad  \textrm{and} \qquad f_1(x\rightarrow 0)=1.
\end{align}
As a consequence of this requirement, it is necessary for the metric potentials $g_{tt}$ and $g_{rr}$ to possess finite values at the central point of the stellar configuration. Furthermore, their derivatives also possess finite values at the center of the star, specifically: $a'(r\rightarrow0)=0$ and $a'_1(r\rightarrow0)=\frac{4c_1k}{R(1+k)^5}$. The mentioned limitations guarantee that the measurement remains consistent at the core and demonstrates a positive characteristic within the inner region of the star.\vspace{0.1cm}\\
ii-At   center of  star, the density (\ref{anz1}) and pressure (\ref{polytrope}) take the following form:
\begin{align}\label{reg1}
\rho(x\rightarrow 0)=\rho_0\, \qquad p(x\rightarrow 0)=k\rho_0{}^2.
\end{align}
By examining Eq.~(\ref{reg1}), when $\rho_0>0$ and $k>0$, it becomes evident that the density and pressure in the central region of the star remain consistently positive. Apart from that, if $\rho_0$ or $k$ is non-positive, the density and pressure can become negative. These observations align with the information depicted in Figures \ref{Fig:1} \subref{fig:pot2}, \ref{Fig:1} \subref{fig:density}, and \ref{Fig:1} \subref{fig:pressure1}, providing further consistency to the discussed aspects. \vspace{0.1cm}\\
iii-The density and pressure gradients of our model are provided in the following manner:
\begin{eqnarray}\label{dsol}
\rho'=-\rho_0, \qquad \qquad p'=-{2k\rho_0{}^2\,(1-x)}\, .
\end{eqnarray}
Here $\rho'=\frac{d\rho}{dx}$ and  $p'=\frac{dp}{dx}$. Equation (\ref{dsol}) illustrates that the components of the energy-momentum   derivatives exhibit negative values.\vspace{0.1cm}\\
iv-The calculation of the speed of sound, using relativistic units where the values of $c$ and $G$ are equal to 1, is achieved by \cite{HERRERA1992206}:
\begin{eqnarray}\label{dso2}
v_r{}^2=\frac{dp}{d\rho}=2\,k \,\rho_0\,(1-x)\, .
\end{eqnarray}

At this point, we are prepared to graphically represent the aforementioned conditions in order to observe their behaviors.
In Figure \ref{Fig:1} \subref{fig:pot2}, we illustrate the characteristics of the metric potentials. As depicted in Figure \ref{Fig:1} \subref{fig:pot2}, the metric potentials take on the values $a_1(x\rightarrow 0)=0.4$ and $a(x\rightarrow 0)=1$ as $x$ approaches 0. This implies that within the central region of the star, both metric potentials exhibit finite positive values.

We proceed to create graphs illustrating the density and pressure, as outlined in Equation (\ref{anz1}), represented in Figure \ref{Fig:1} \subref{fig:density} and \ref{Fig:1} \subref{fig:pressure1}.  As depicted in Figure \ref{Fig:1} \subref{fig:density} and \ref{Fig:1} \subref{fig:pressure1}, the components of the energy-momentum exhibit positive values, which are consistent with predictions for a reasonable stellar arrangement. Moreover, as depicted in Figure \ref{Fig:1} \subref{fig:density} and \subref{fig:pressure1}, the components of the energy-momentum tensor exhibit elevated values at the core and gradually decrease as they approach the periphery. These observed patterns are characteristic of a plausible star.
\begin{figure}
\centering
\subfigure[~Gradient of density]{\label{fig:drho}\includegraphics[scale=0.3]{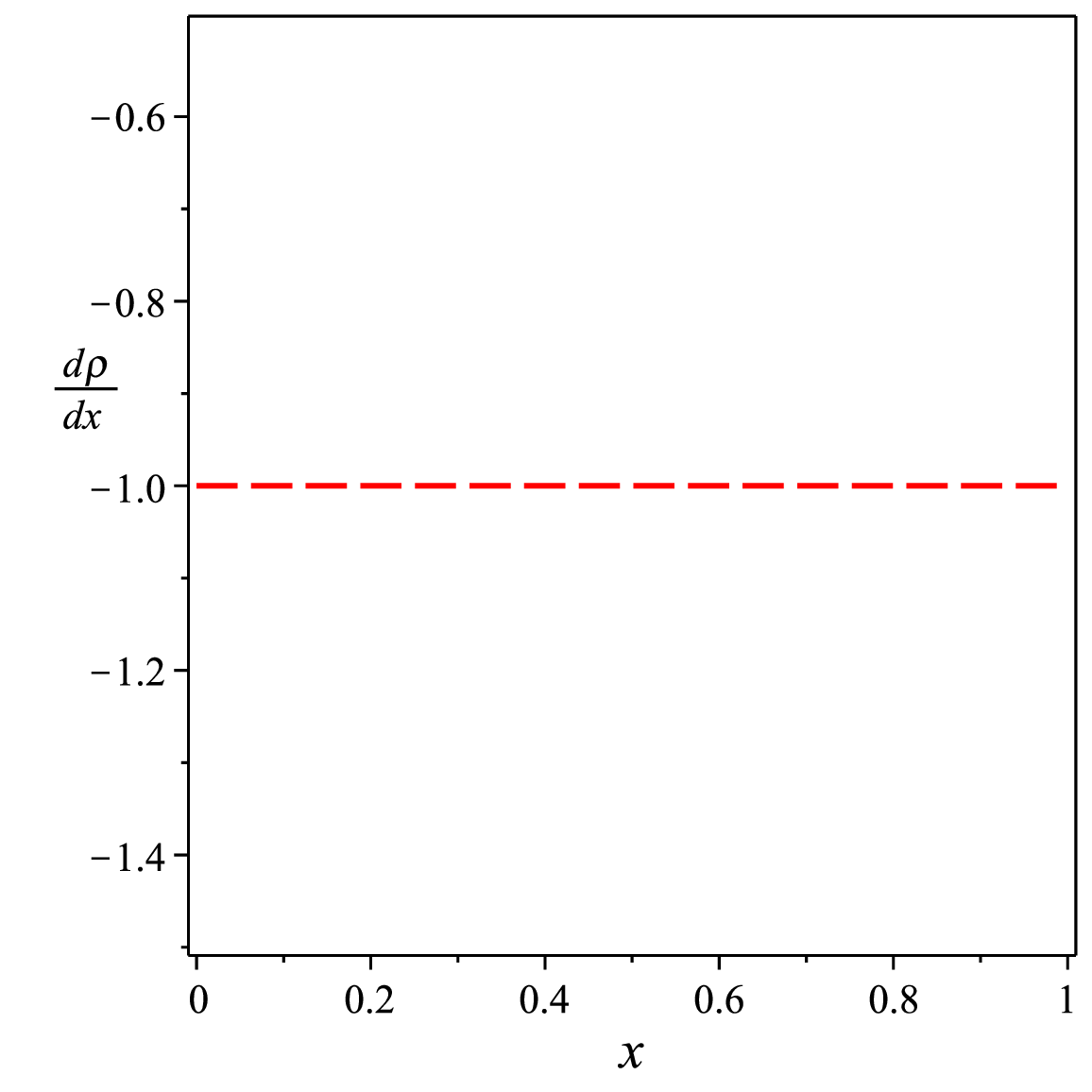}}
\subfigure[~Gradient of pressure]{\label{fig:dp}\includegraphics[scale=.3]{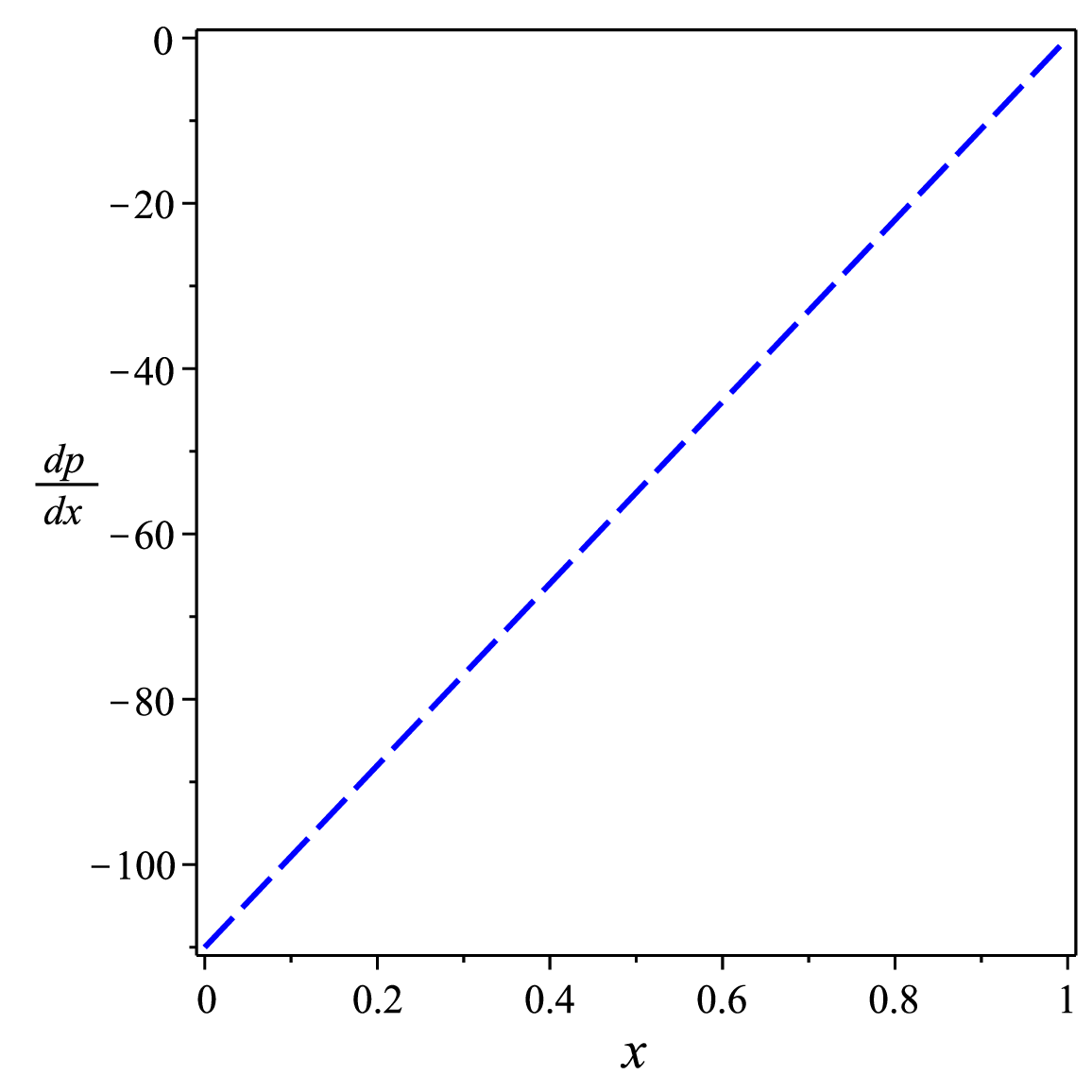}}
\caption[figtopcap]{\small{{{ A graph is shown, displaying the variations in the gradient of density and pressure from Equation (\ref{anz1}), plotted against the dimensionless value of $x$.}}}}
\label{Fig:2}
\end{figure}

Figure \ref{Fig:2} illustrates the presence of adverse values in the derivatives of the components of the energy-momentum tensor, indicating a uniform decrease in both density and pressure throughout the entirety of the star's structure.
\begin{figure}
\centering
\subfigure[~Sound speed]{\label{fig:dpr}\includegraphics[scale=0.3]{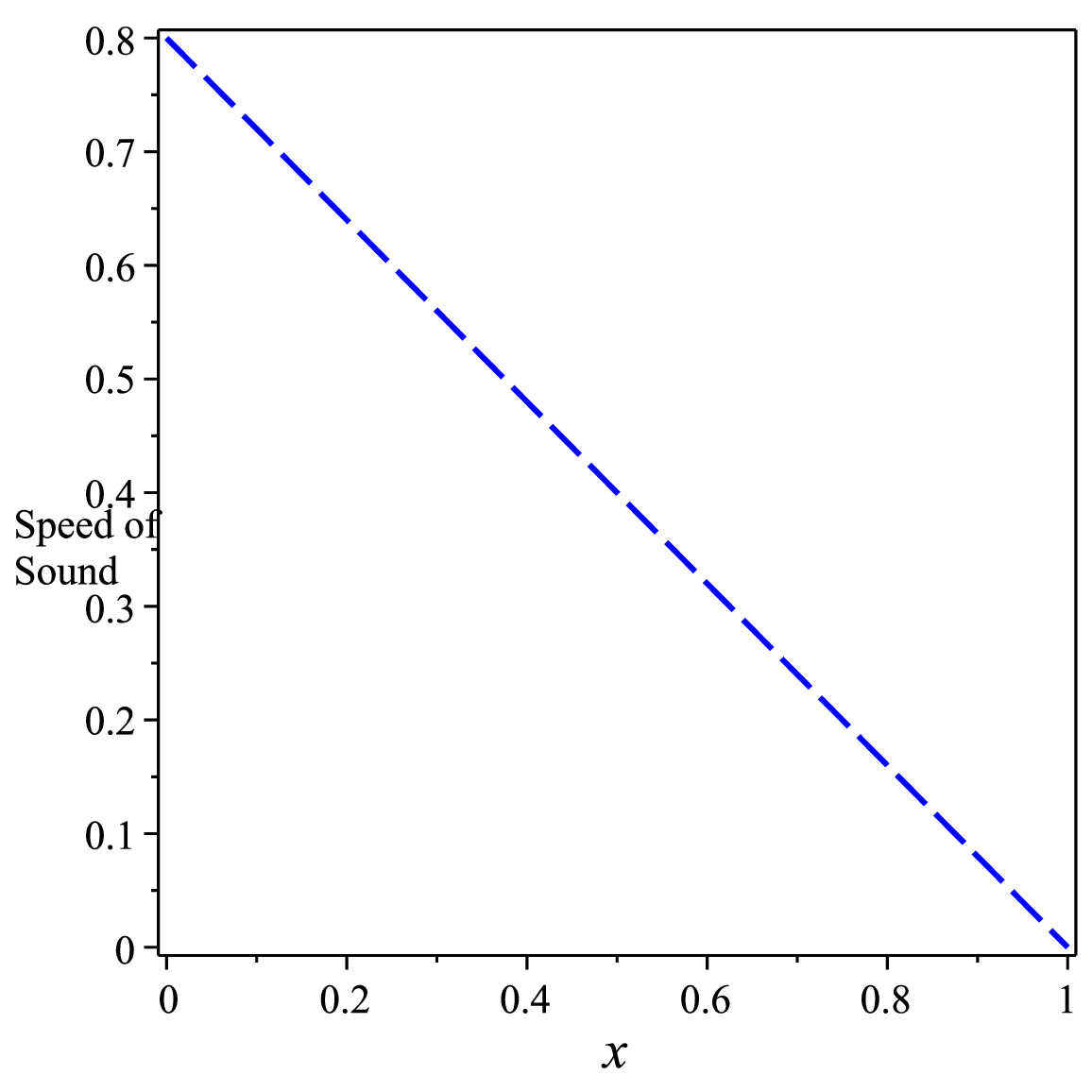}}
\subfigure[~Relationship between mass and radius]{\label{fig:pressure}\includegraphics[scale=.3]{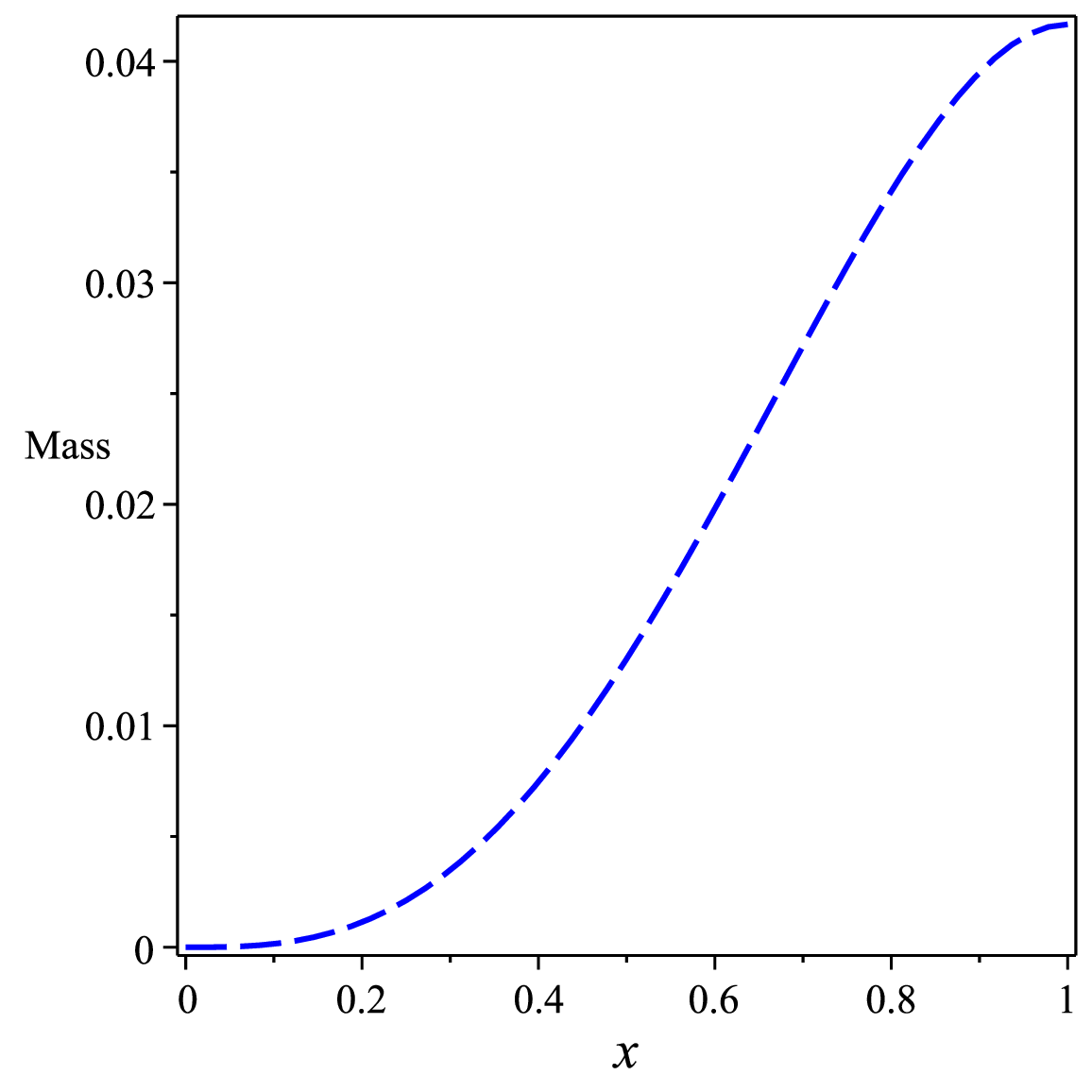}}
\subfigure[~The compactness]{\label{fig:comp}\includegraphics[scale=.3]{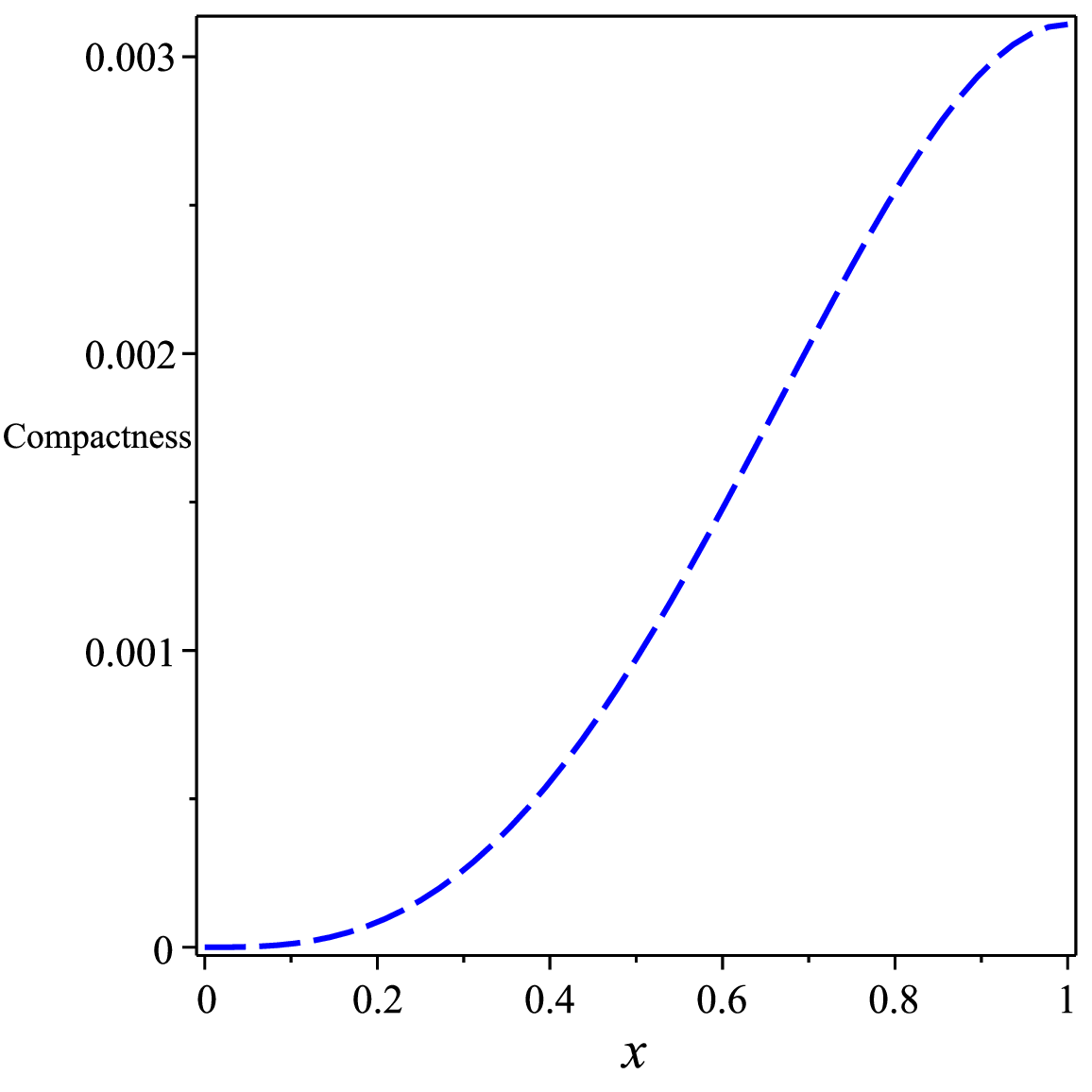}}
\caption[figtopcap]{\small{{{A diagram is presented, depicting the behavior of various quantities with respect to the dimensionless value of $x$. Specifically, the plot includes the sound speed (labeled as \subref{fig:dpr}), the relation between mass and radius (labeled as \subref{fig:pressure}), and the level of compactness exhibited by the celestial body(labeled as \subref{fig:comp}).}}}}
\label{Fig:3}
\end{figure}
Figure \ref{Fig:3} presents visual depictions of the speed of sound, the relationship between mass and radius, and the compactness parameter. As depicted in Fig. \ref{Fig:3}\subref{fig:dpr}, the speed of sound is found to be less than one, which verifies that the causality condition is not violated within the interior of the stellar configuration when the parameter of the equation of state (EoS) is $k<0.5$.  Furthermore, Fig. \ref{Fig:3}\subref{fig:comp} illustrates that the compactness of our model is restricted within the interval of $0<C<0.003$, where $C$ is defined as the ratio of $M$ to $xR$ in the stellar arrangement.
\begin{figure}
\centering
\subfigure[~Null  energy conditions]{\label{fig:NEC}\includegraphics[scale=0.3]{JFRMMM_Danial_den.eps}}
\subfigure[~Weak  energy conditions]{\label{fig:WEC}\includegraphics[scale=0.3]{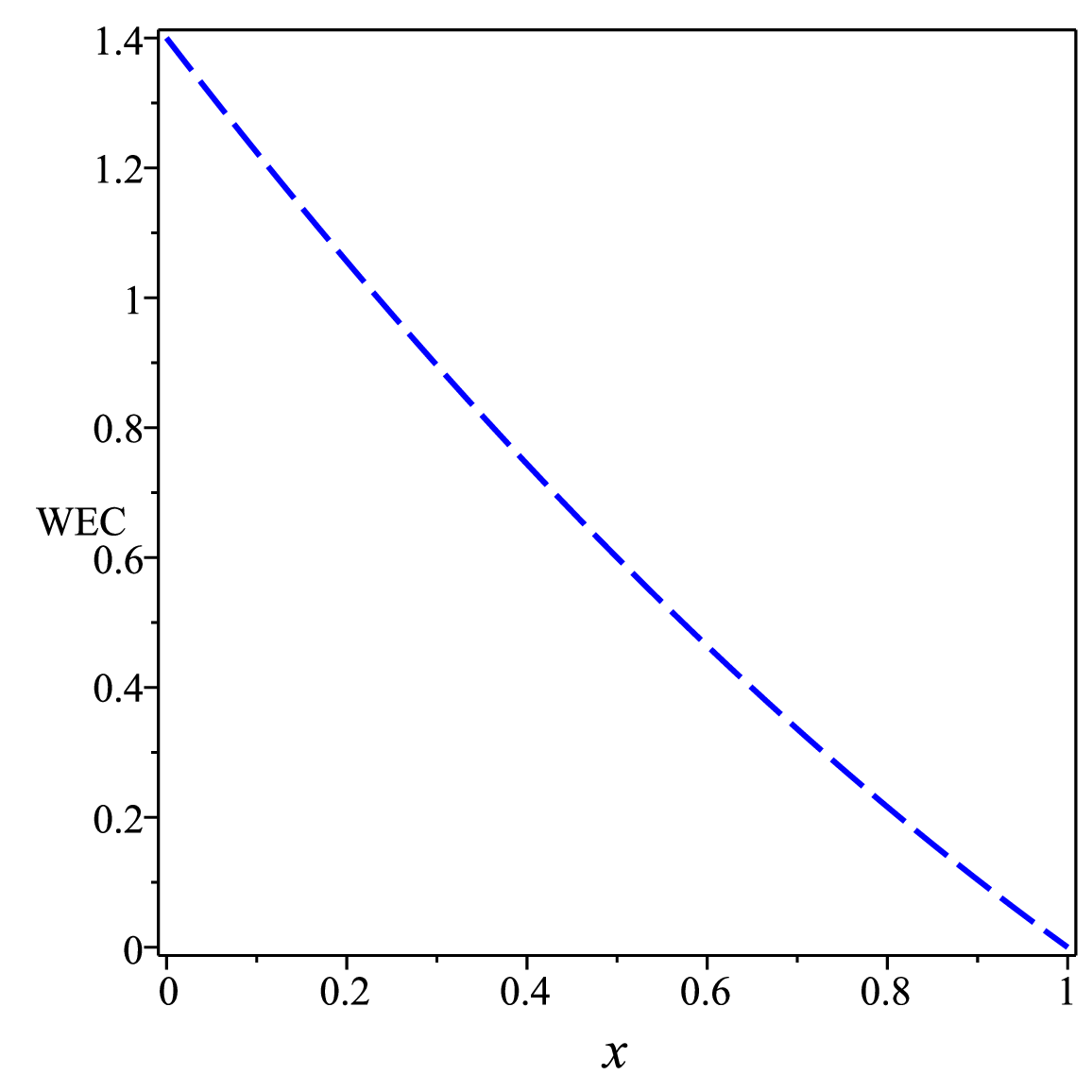}}
\subfigure[~Strong energy condition]{\label{fig:SEC}\includegraphics[scale=.3]{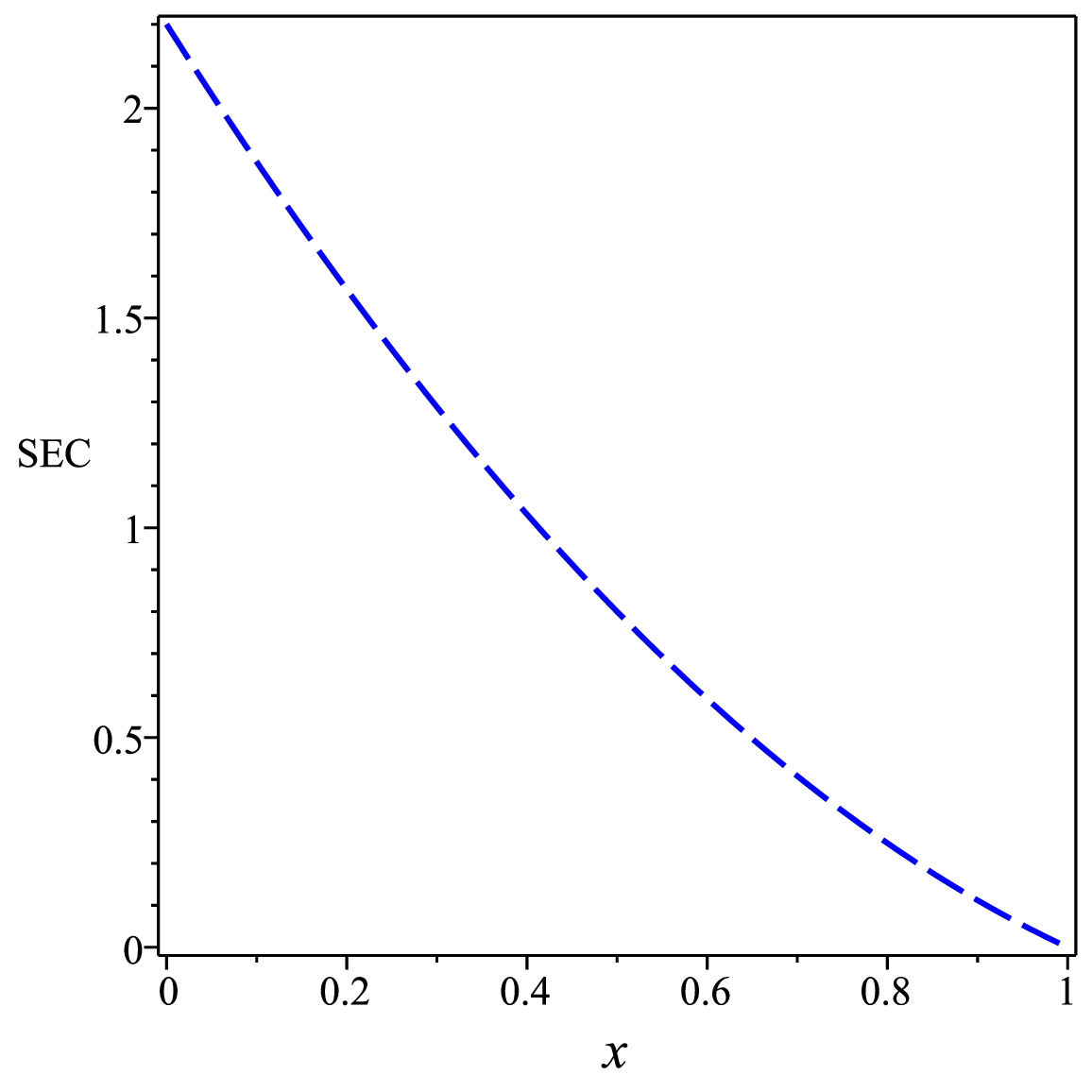}}

\caption[figtopcap]{\small{{{A graph is generated to illustrate the behavior of the null, weak, and strong energy conditions, as determined by Equation (\ref{anz1}), plotted against the dimensionless value of $x$.}}}}
\label{Fig:5}
\end{figure}

The energy conditions are depicted in Figure \ref{Fig:5}, showcasing the characteristics of each condition.  More specifically, Fig.~\ref{Fig:5} \subref{fig:NEC}, \subref{fig:WEC}, and \subref{fig:SEC} display the presence of positive values for the NEC (Null Energy Condition), WEC (Weak Energy Condition), and SEC (Strong Energy Condition) respectively. This verification provides assurance that all energy conditions are met across the entire stellar configuration, aligning with the criteria for a physically feasible stellar model.
\begin{figure}
\centering
\subfigure{\label{fig:EoS}\includegraphics[scale=0.4]{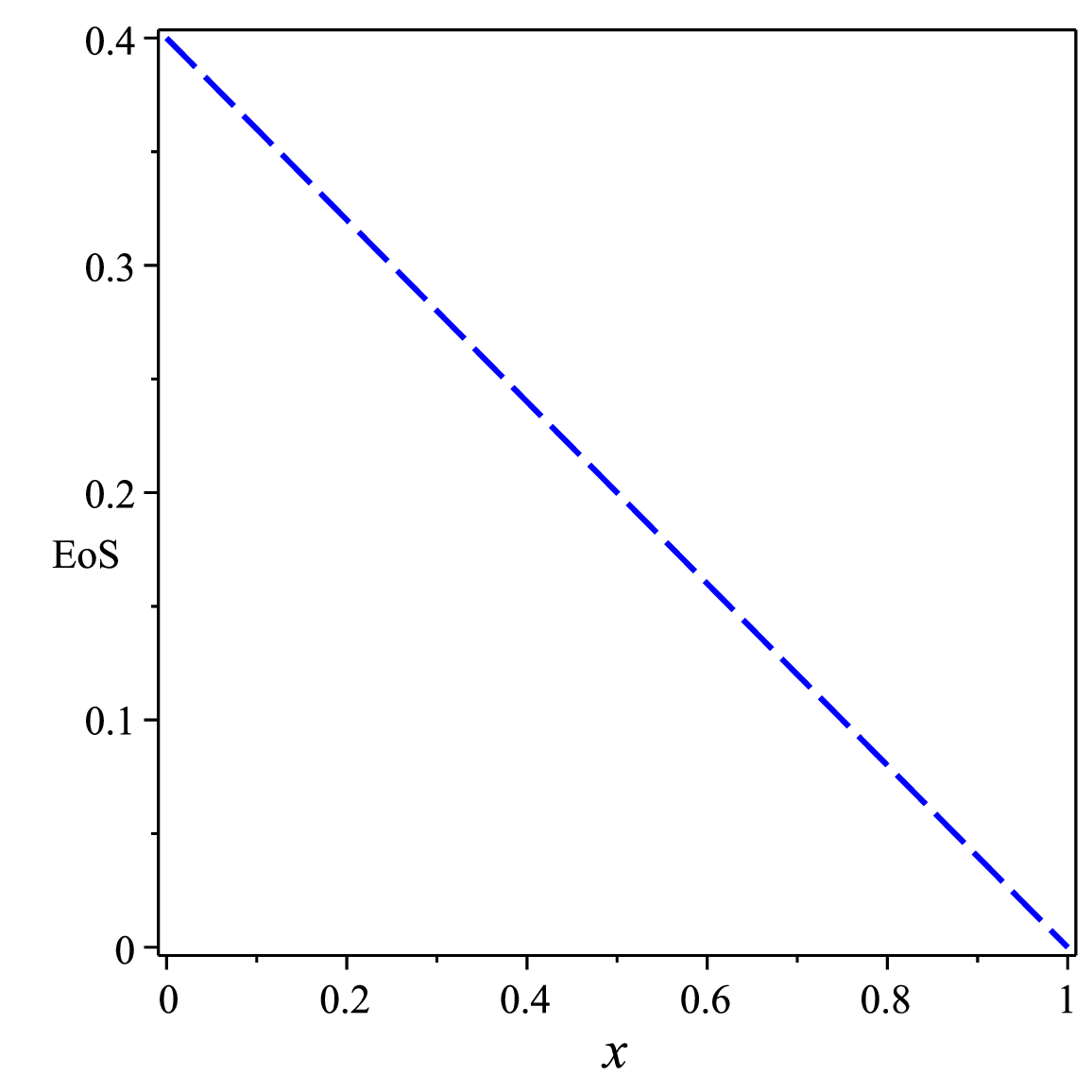}}
\subfigure{\label{fig:Z}\includegraphics[scale=.4]{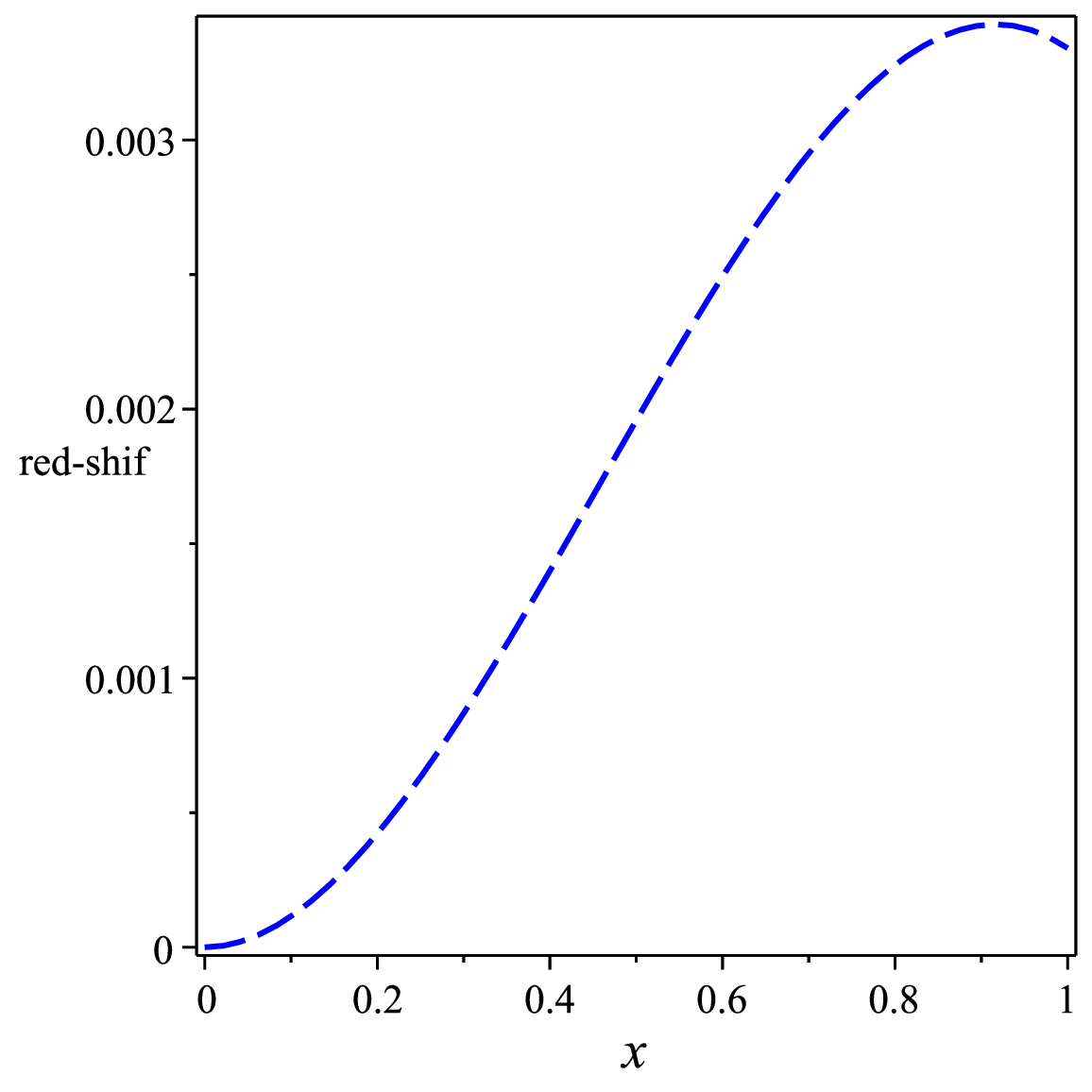}}
\caption[figtopcap]{\small{{A plot is presented to visualize the equation of state (EoS) as a function of the radial coordinate $r$, labeled as \subref{fig:EoS}. Additionally, the redshift is plotted and labeled as \subref{fig:Z}.}}}
\label{Fig:6}
\end{figure}

Figure \ref{Fig:6} represents the plot of the Equation of State (EoS). In particular, Figure \ref{Fig:6} \subref{fig:EoS} indicates that the EoS exhibits a linear behavior.
\section{The model's stability}\label{stability}
At this point, we are prepared to examine the stability concern of our model by conducting tests involving the the index of adiabatic and the static case.
\begin{figure}
\centering
{\label{fig:gamma}\includegraphics[scale=0.3]{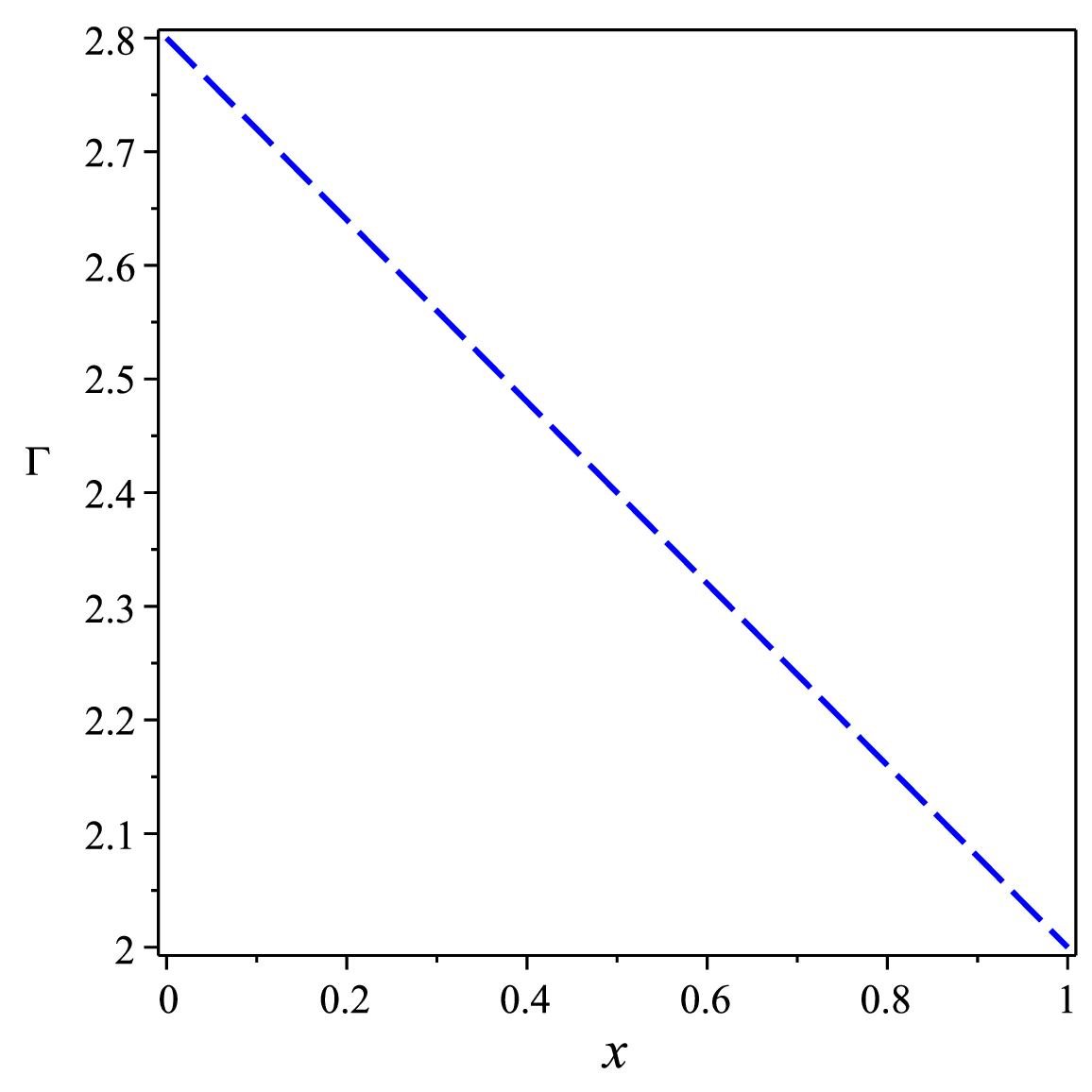}}
\caption[figtopcap]{\small{{Plot of the gravitational,  and the hydrostatic  forces  vs. the dimensionless $x$.}}}
\label{Fig:7}
\end{figure}
\subsection{An adiabatic index}
To investigate the stable balance of a spacetime that possesses spherical symmetry, an analysis of the adiabatic index can be conducted. The adiabatic index plays a vital role in evaluating the stability requirement, serving as an essential instrument for this purpose.   Specifically, the adiabatic perturbation, denoted as $\Gamma$, is defined as follows \cite{1964ApJ...140..417C,1989A&A...221....4M,10.1093/mnras/265.3.533,Nashed:2011fg,Nashed:2021sji,Nashed:2011fg,Nashed:2020kjh,Roupas:2020mvs}:
\begin{eqnarray}\label{ai}  \Gamma=\left(\frac{\rho+p(x)}{p(x)}\right)\left(\frac{dp(x)}{d\rho(x)}\right)\,.
 \end{eqnarray}
 If the adiabatic index $\Gamma$ is greater than $\frac{4}{3}$, a Newtonian isotropic sphere will have a stable equilibrium
 \cite{1975A&A....38...51H}.

 Using Eq. (\ref{ai}),  we get
 \begin{eqnarray}\label{a12}  &&\Gamma=2+2k \rho_0(1-x).
 \end{eqnarray}
Figure \ref{Fig:7} displays the adiabatic index $\Gamma$. The graph clearly indicates that the value of $\Gamma$ remains consistently above $4/3$ throughout the star's interior. Therefore, we can infer that the stability requirement is met.

\subsection{Stability in the stationary state}
Another approach to validate the stability of model (\ref{anz1}) involves investigating the static state proposed by Harrison, Zeldovich, and Novikov \cite{1965gtgc.book.....H,1971reas.book.....Z,1983reas.book.....Z}.  In a previous study by Harrison, Zeldovich, and Novikov, it was established that a stable configuration of a star necessitates a positive and increasing derivative of mass with respect to the central density $\rho(x \rightarrow 0)$, denoted as $\frac{\partial M}{\partial \rho_0} > 0$. By applying this condition, we can determine the specific form of the central density as follows:
\begin{eqnarray}\label{rm}
\rho(r\to 0)=\rho_0\,.
\end{eqnarray}
Subsequently, utilizing Eq. (\ref{MRs}), we can derive the mass corresponding to the central density, denoted as:
\begin{eqnarray}
 \label{rm11}
&&M(\rho_0)4\pi \int_0^{R} y^2 \rho(y) dy
= \frac{\pi \rho_0 R^3}{3}\,.
\end{eqnarray}
The behavior of the mass derivative with respect to the central density can be described by the following pattern:
\begin{eqnarray}
 \label{rm111}
&&\frac{\partial M(\rho_0)}{\partial \rho_0}=\frac{\pi  R^3}{3}\,.
\end{eqnarray}
Equations  (\ref{rm11}) and (\ref{rm111}) guarantee the verification of
stability condition of our  model.
\section{Discussion and conclusions}\label{S5}


The primary objective of this investigation is to analyze a compact star's static configuration, which possesses both spherical symmetry and is within the framework of mimetic gravity theory. This theory incorporates two main components: a scalar field and a Lagrangian multiplier.  In our formulation, we demonstrated the ability to construct a model that accurately replicates the profile of a given spherically symmetric spacetime, regardless of the equation of state (EoS) for matter and energy density.
The shape of the density function $\rho(x)$ is of utmost importance in determining the radius $R$ and mass $M$ of the compact star. By manipulating the Lagrangian multiplier $\lambda(\eta)$, it becomes feasible to establish a flexible connection between the radius $R$ and the compact star. This creates a situation where the Lagrangian multiplier $\lambda(x)$ and the equation of state (EoS) describing the model exhibit a degenerate relationship. As a result, it is clear that relying solely on the mass-radius relation is inadequate for fully constraining the model.

To illustrate this further, we take a closer look at the polytrope equation of state (EoS) given by (\ref{polytrope}). By selecting a specific form for the  density in Eq.~ (\ref{anz1}), we construct a practical isotropic model.   We then proceed to investigate the physical characteristics of this model. Through rigorous analysis using different analytical techniques and validations, we carefully scrutinize the obtained analytic solution. This comprehensive examination enables us to observe and analyze the physical behavior manifested by our solution.

It is important to highlight that the preceding discussion confirms the satisfaction of all the physical conditions by the spherically symmetric interior spacetime configuration considered in this study within the framework of mimetic gravitational theory coupled with a Lagrangian multiplier. However, it should be noted that an alternative form of mimetic gravitational theory that does not involve the coupling with a Lagrangian multiplier may not support the existence of the isotropic model, as evidenced by equations (\ref{eqtt}), (\ref{eqrr}), and (\ref{fe}).

{ Moreover, the field equations governing the equilibrium of rapidly rotating neutron
stars in scalar-tensor theories of gravity, as well as representative numerical solutions are discussed  in \cite{Doneva:2013qva}. New models of slowly rotating, perfect-fluid neutron stars are constructed  by extending the classical
Hartle-Thorne formalism to generic scalar-tensor theories of gravity \cite{Pani:2014jra}. An investigation of self-consistently slowly rotating neutron and strange stars in R-squared gravity is investigated in \cite{Staykov:2014mwa}. A study of static neutron stars in the framework  of a class of non-minimally coupled inflationary potentials have been presented in \cite{Oikonomou:2021iid}. Are all the previous studies presented in \cite{Doneva:2013qva,Pani:2014jra,Staykov:2014mwa,Oikonomou:2021iid} can be discussed in the framework of the present study? This will be done elsewhere.}


%
\end{document}